\documentclass{ws-procs9x6}

\begin{document}

\title{Revealing the Dark TeV Sky: The Atmospheric Cherenkov Imaging Technique
for Very High Energy Gamma-ray Astronomy
\footnote{\uppercase{T}his work is respectfully dedicated
to the memory of \uppercase{N}eil \uppercase{A}. \uppercase{P}orter (1930-2006), 
one of the \uppercase{F}ounding \uppercase{F}athers of
\uppercase{V}ery \uppercase{H}igh \uppercase{E}nergy 
\uppercase{G}amma-ray \uppercase{A}stronomy.}}

\author{Trevor C. Weekes}

\address{   Whipple Observatory,\\
 Harvard-Smithsonian Center for Astrophysics,\\
            P.O. Box 6369, Amado, Arizona 85645-0097, U.S.A.\\
e-mail:tweekes@cfa.harvard.edu}

\maketitle

\abstracts{
The Atmospheric Cherenkov Imaging Technique has opened up the gamma-ray
spectrum from 100 GeV to 50 TeV to astrophysical exploration. The
development of the technique (with emphasis on the early days)
 is described as are the basic principles
underlying its application to gamma-ray astronomy. 
The current generation of arrays of telescopes, in particular, VERITAS
 is briefly described.
}

\section{Introduction}

One of the last frontiers of the gamma-ray sky is that characterized by
the distribution of
TeV photons. These photons can be detected relatively easily with ground-based
detectors (constituting a TeV ``window'' in the atmosphere); thus the
detection of TeV gamma-ray sources did not have
to await the availability of space platforms. In practice although the
technology
was available
at an early date, it required the impetus of gamma-ray space astronomy to
justify a major
effort in a new discipline. Since it concerns the highest energy photons
with which
it is yet
feasible to map the sky, it is of particular interest to high energy
astrophysicists.
Any source of TeV photons must be associated with a cosmic
particle accelerator and of inherent interest to high energy particle
physicists as well as
students of the cosmic radiation.

To date almost all the observational results in the energy interval
100 GeV - 100 TeV have come from observations using the so-called
``Atmospheric Cherenkov
Imaging Technique (ACIT).''  Although considerable effort has been applied
to the development of alternative techniques, they are more specialized 
and will not be considered here.

In this historical review of the ACIT, emphasis will be on the early days
in which the technique was established; a brief outline of the general 
principles underlying atmospheric Cherenkov telescopes (ACT) will be given
and a description, albeit incomplete, of the
ACIT as currently used and the present generation of instruments will be
described.
More complete accounts can be found elsewhere\cite{porter1981},\cite{weekes1988},
\cite{Aharonian:97},\cite{fegan1997},\cite{ong},\cite{weekes2003}.

\section{Early History of the Atmospheric Cherenkov Technique}

\subsection{Discovery of the Phenomenon}

\begin{figure}
\begin{center}
\includegraphics[width=.4\textwidth]{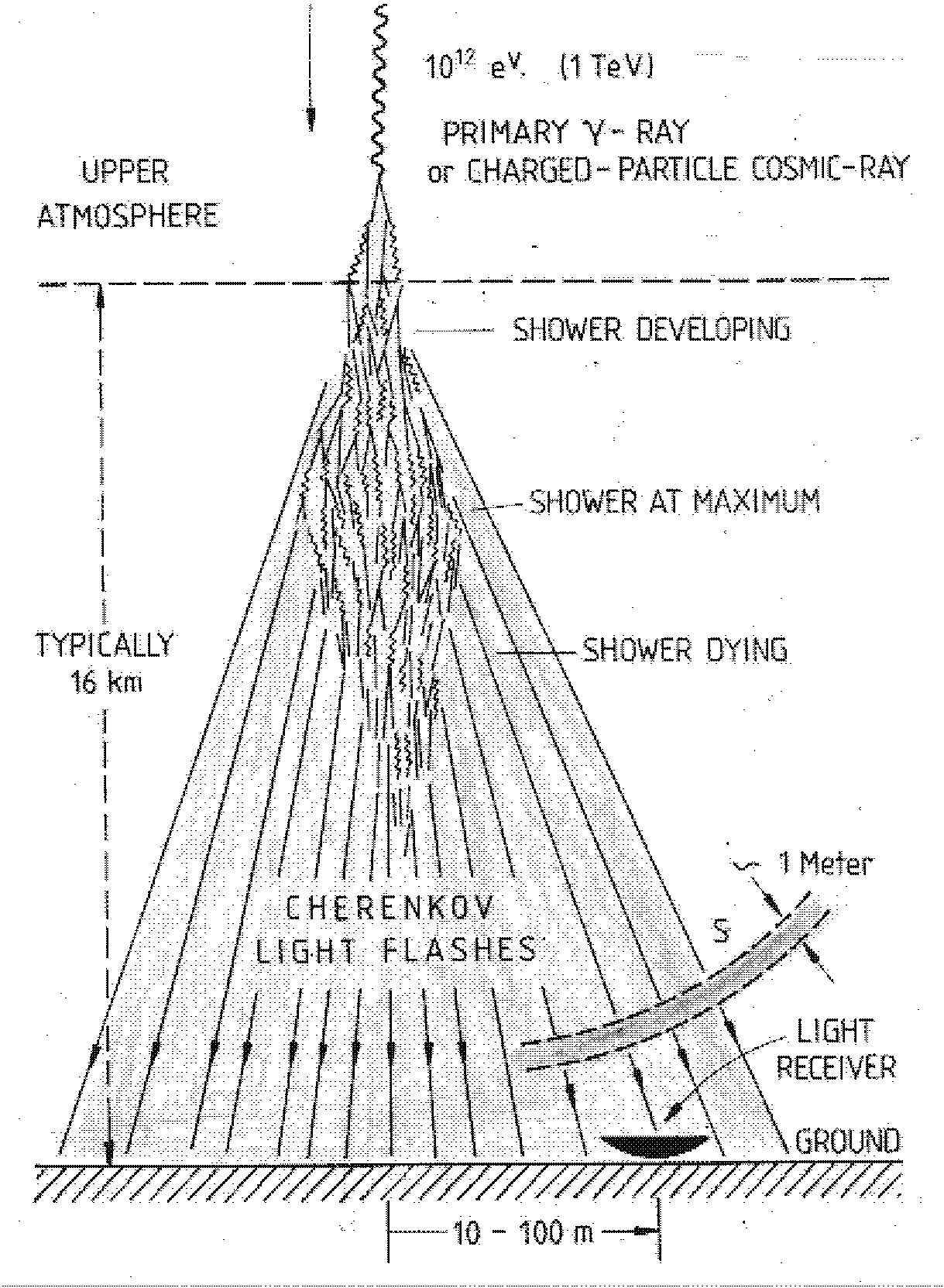}
\hspace{1cm}
\includegraphics[width=.4\textwidth]{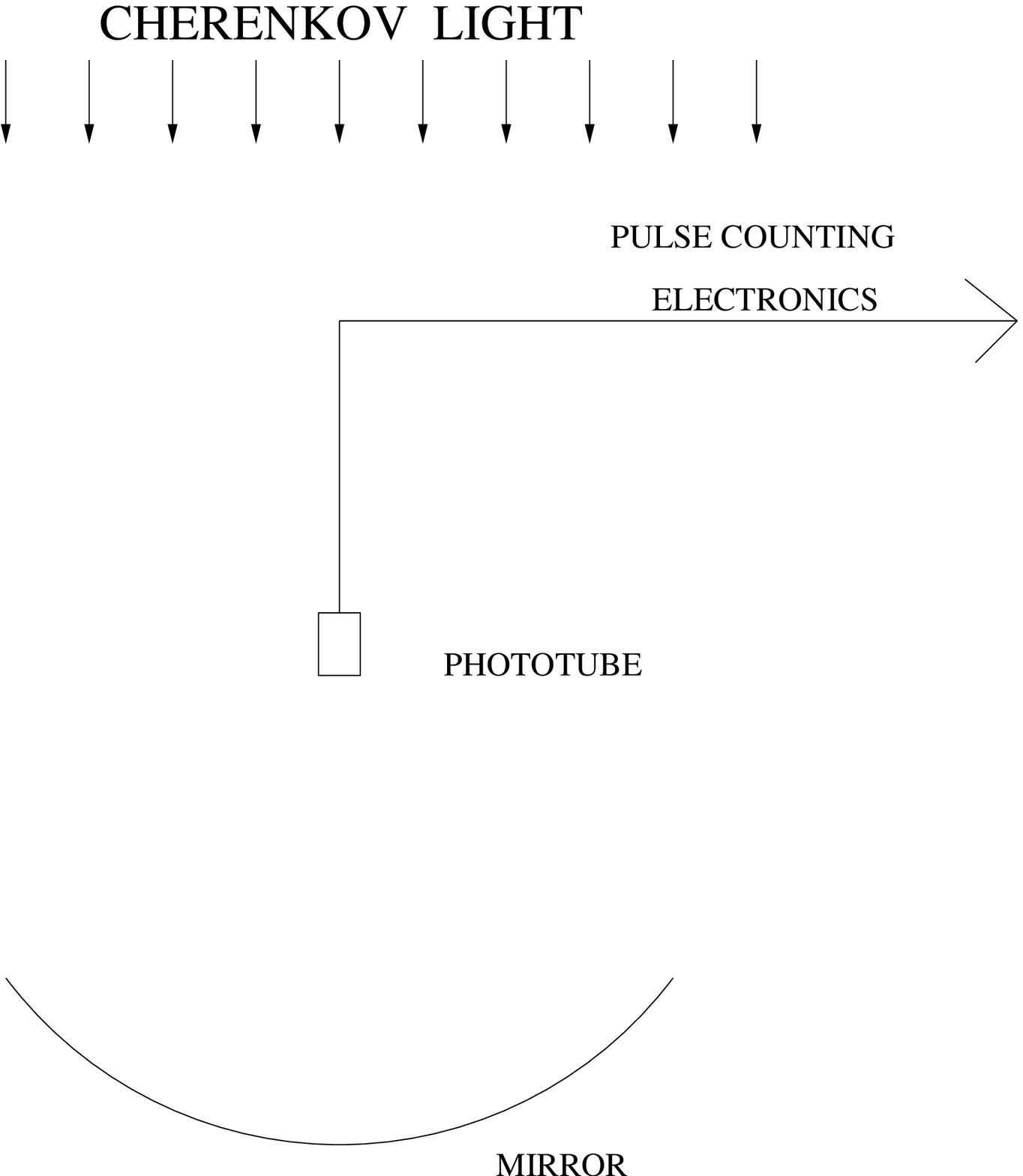}
\end{center}
\caption[]{{\bf Left:} Cartoon of the atmospheric Cherenkov shower 
phenomenon, as drawn by John Jelley in 1993. {\bf Right:} The essential 
elements of an Atmospheric Cherenkov Detector}
\label{jvj}
\end{figure}

In the Ph.D. dissertations of students studying the atmospheric Cherenkov phenomenon,
the first reference is usually to the 1948 note in the Royal Society report on 
the study of night-sky light and aurora by the British Nobel Laureate, P.M.S.
Blackett\cite{blackett}; in that note
he points out that perhaps 0.01\% of the light in the dark night-sky must come from
Cherenkov light emitted by cosmic rays and their secondary components as they
traverse the atmosphere.
Little attention was paid to this prediction (since it seemed unobservable)
at the time. Fortunately five years later, when Blackett was visiting the Harwell Air Shower
array, he brought his prediction to the attention of two Atomic Energy Research
Establishment physicists, Bill Galbraith and John Jelley. After the visit, the idea
occurred to them that, while the net flux of Cherenkov light would be impossible to
measure, it might just be possible to detect a short light pulse from a cosmic ray air shower
which involved some millions of charged particles (Figure~\ref{jvj}).

Within a week Galbraith and Jelley had assembled the items necessary to test their
hypothesis. A 5 cm diameter photomultiplier tube (PMT)
was mounted in the focal plane of a 25
cm parabolic mirror (all housed in a standard-issue Harwell garbage can) and
coupled to an amplifier with a state-of-the-art 5 MHz amplifier whose output was
displayed on an oscilloscope. They observed oscilloscope triggers
from light pulses that exceeded the average
noise level of the night-sky background every two minutes. They noted that the
pulses disappeared when the garbage can lid was put in place and a padding lamp
was adjusted to give the same current in the PMT as was observed from the 
night-sky\cite{gj1952}.
Jelley noted that if the rate had been any lower than that observed they would
probably have given up and gone home early!\cite{jvj1993}. It is not often that a new
phenomenon can be discovered with such simple equipment and in such a short
time, but it may also be true that it is not often that one finds experimental
physicists with this adventurous spirit! Whereas the modern physicist would not
embark on a speculative venture of this nature without extensive simulations, 
John Jelley had a great suspicion of excessive computation and relied instead
on his gut feelings for the inherent physics of the phenomenon; 
he was seldom wrong!

\subsection{The Power of the Technique}

With the Harwell air shower array (one of the largest
such arrays then in existence) in close
proximity, it was easy to show that the light pulses
were indeed associated with air showers.
In the years that followed, Galbraith and Jelley  made a series of
experiments in which they determined the basic parameters of the Cherenkov
radiation from air showers. The account of these elegant experiments is a must-read
for all newcomers to the field\cite{gj1953},\cite{jg1953}.
The basic detector elements of the ACT are extremely simple (Figure~\ref{jvj}).
It was realized at an early stage that
the phenomenon offered the possibility of detecting point sources of cosmic
ray air showers with high efficiency. Since charged primaries are rendered
isotropic by the intervening interstellar magnetic fields, in practice this meant
the detection of point sources of neutral quanta, i.e., gamma-ray photons or
perhaps neutrons.
The lateral spread of the Cherenkov light from the shower
as it strikes the ground is $\approx$ 100-200 m
so that even a simple light receiver of modest dimensions has an effective collection
area of some tens of thousands of square meters. The fact that the light pulse
preserves much of the original direction of the primary particle and that the
intensity of light is proportional to the total number of secondary particles,
and hence to the energy of the primary, makes the detection technique potentially
very powerful.

\begin{figure}
\begin{center}
\includegraphics[width=.6\textwidth]{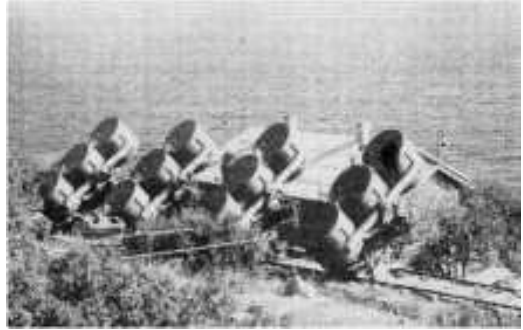}
\end{center}
\caption[]{The first ground-based experiment in TeV gamma-ray astronomy which 
was the Lebedev Institutes's of twelve 1.5 m searchlight mirrors in the Crimea; 
it had an energy threshold of 1.5 TeV}
\label{crimea}
\end{figure}

\begin{figure}
\begin{center}
\includegraphics[width=.3\textwidth]{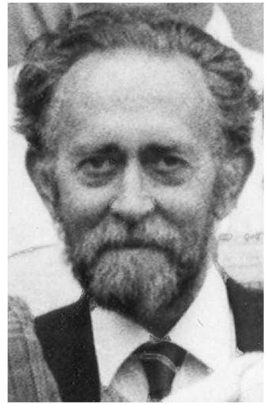}
\hspace{1cm}
\includegraphics[width=.4\textwidth]{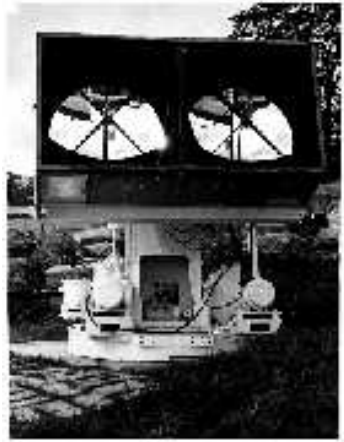}
\end{center}
\caption[]{{\bf Left:} Neil A. Porter (1930-2006) (Photo: D.J.Fegan)  
{\bf Right:} The second ground-based gamma-ray telescope; the 
British-Irish experiment at Glencullen,
Ireland c. 1964; the telescope
consisted of two 90 cm searchlight mirrors on a Bofors gun mounting.
The experiment was led by Jelley and Porter.}
\label{glencullen}
\end{figure}

\begin{figure}
\begin{center}
\includegraphics[width=.7\textwidth]{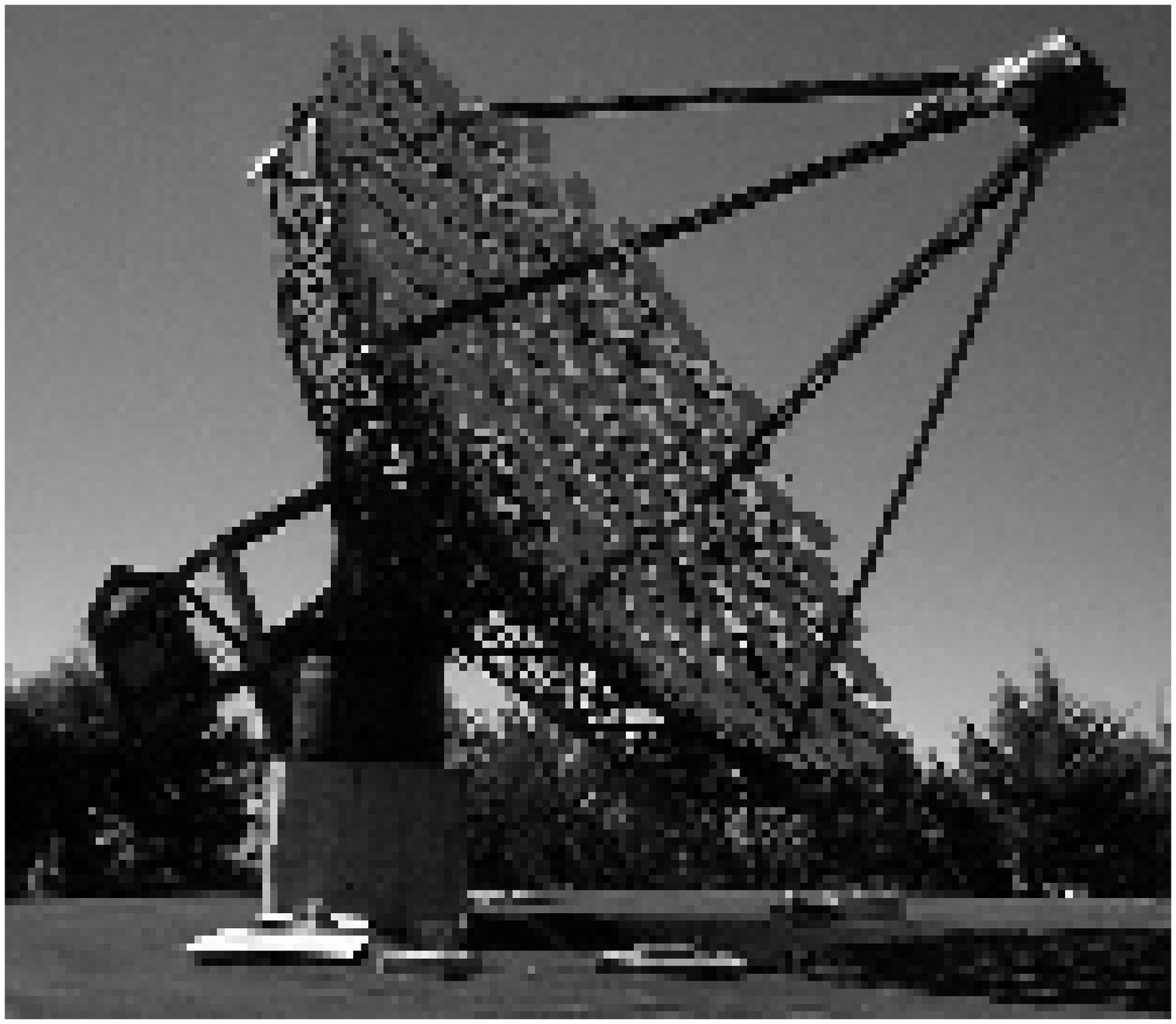}
\end{center}
\caption[]{The Whipple Observatory 10 m gamma-ray telescope was built in 1968;
it is still in operation. It is composed of 250 glass facets, each of focal
length 7.3 m.}
\label{whipple}
\end{figure}

The prediction by Cocconi\cite{cocconi1959} of a strong flux of
TeV gamma rays from the Crab
Nebula precipitated an experiment by the Lebedev Research Institute in the Crimea
in 1960-64\cite{chudakov1965}.  Supernova
Remnants and Radio Galaxies had recently been identified as sources containing
synchrotron-emitting electrons which suggested that they might be gamma-ray sources.
A selection of these (including the Crab Nebula) were examined with a ACT
system consisting of twelve 1.5 m aperture ex-World War II 
searchlight mirrors mounted on railway cars at a dark site 
near the Black Sea (Figure~\ref{crimea}). This system did not
attempt to discriminate between air showers initiated by gamma rays and those
initiated by hadrons. No sources were found but the basic methodology
involved in a search for point source anisotropies in the cosmic ray air shower
distribution was defined. The technique was refined by John Jelley and Neil Porter
in a pioneering British-Irish experiment in the Dublin Mountains
 in which the candidate source list was expanded to include the recently discovered
quasars and magnetic variable stars (with null results \cite{long1964}). This early
experiment also used ex-World War II searchlight mirrors on a Bofors gun mounting
(continuing the tradition of putting military hardware to good use) 
(Figure~\ref{glencullen}). The Smithsonian group led by Giovanni Fazio built the
first large optical reflector for gamma-ray astronomy 
on Mount Hopkins in southern Arizona (Figure~\ref{whipple}). This 10 m telescope is still
in use after 38 years of service! This again was a first generation device in
which the assumption was made that there was no easily measured differences in the
light pulses from gamma-ray and hadronic primaries. The motivation for this large
increase in mirror area (and decrease in energy threshold) was a refined prediction
of a detectable flux of gamma rays from the Crab Nebula based on a
Compton-synchrotron model\cite{gould1964}.

Although these first generation detection systems were extremely simple and
exploited the ease with which gamma rays could be {\it detected}, they did not
provide the means of {\it identifying} gamma rays among the much more numerous
cosmic ray background.
Hence, until 1989 when the Crab Nebula was finally detected\cite{weekes1989}, 
there was no credible detection of a
gamma-ray flux from any cosmic source.

\subsection{Basic Principles}

The light signal (in photoelectrons) detected is
given by:\\
S = $\int^{\lambda_1}_{\lambda_2}$ k E($\lambda$)  T($\lambda$) $\eta(\lambda$) A
d$\lambda$ \\
where C($\lambda$) is the Cherenkov photon flux within the
wavelength sensitivity
bounds of the PMT, $\lambda_1$ and $\lambda_2$,
E($\lambda$) is the shower Cherenkov emission spectrum
(proportional to 1/$\lambda^2$), T($\lambda$) is the atmospheric
transmission  and k is a constant which
depends on the shower, and the geometry.

 The signal must be detected above the
fluctuations in the night-sky background during
the integration time of the pulse counting system, $\tau$.

The sky noise B is given by:\\
B = $\int^{\lambda_1}_{\lambda_2}$ B($\lambda$) $\eta$($\lambda$)
$\tau$ A $\Omega$ d$\lambda$.\\
Hence the signal-to-noise ratio is essentially\\
S/N = S/B$^{0.5}$  = $\int^{\lambda_1}_{\lambda_2}$ C($\lambda$)
[$\eta$
($\lambda$) A /$\Omega$ B($\lambda$) $\tau$]$^{1/2}$ d$\lambda$.

The smallest detectable light pulse is inversely proportional to
S/N; the minimum detectable gamma ray then has an energy threshold,
E$_T$ given by\\
E$_T$ $\propto$ 1/C($\lambda$) [B($\lambda$) $\Omega$
$\tau$/{$\eta$($\lambda$) A]$^{1/2}$

If S = the number of gamma rays detected from a given source in a
time, t,
and  A$_\gamma$ is the collection area for gamma-ray detection,
then
S = F$_\gamma$(E) A$_\gamma$ t.
The telescope will register a background, B, given
by:\\
B = F$_{cr}$ A$_{cr}$(E) $\Omega$ t,
where A$_{cr}$(E) is the collection area for the detection of
cosmic rays of energy E.
The cosmic ray background has a
power law spectrum:\\
F$_{cr}$($>$E) $\propto$ E$^{-1.7}$ and if we
assume the gamma-ray source  has the form:\
F$_\gamma$($>$E$_\gamma$) $\propto$ E$_\gamma$$^{-a_\gamma}$.

Then the standard deviation,
$\sigma$ $\propto$ S/B$^{1/2}$    $\propto$ E$^{1.7/2-a_\gamma}$
[A$_\gamma$/A$_{cr}^{1/2}]$ t$^{1/2}$

The minimum number of standard deviations, $\sigma$, for a reliable
source detection is generally taken as 5\cite{weekes2003}.

\section{Early Development of the ACIT}

\subsection{Discrimination Methods}

At an early stage it was realized that while the atmospheric Cherenkov technique
 provided a very easy way of {\it detecting}
gamma rays with simple light detectors, it did not
readily provide a method of discriminating the light pulse from gamma-ray air showers
from the background of light pulses from the much more numerous cosmic ray showers;
thus the {\it flux sensitivity} was severely limited.
Although the hadron showers are isotropic, 
there is typically a ratio of  1,000-10,000 of cosmic
rays to gamma rays recorded by the simple light detectors that were
available in the two decades following the Harwell experiments.
Once it was apparent that the early, very optimistic, predictions of the strength
of the most obvious potential TeV sources were not to be realized, then attention turned to
methods of improving the flux sensitivity of the technique. Although superficially very
similar, Monte Carlo simulations of shower development and Cherenkov light emission
suggested some differences that might be exploited to preferentially select gamma rays.

These differences are listed below: 
\begin{itemize}
\item Lateral Spread at ground level: the light pool from gamma-ray showers is more
uniform than that from cosmic ray showers. This feature is difficult to exploit since it
requires numerous light detectors spread over relatively large areas; it has recently been
used by the group at the Tata Institute at their Pachmari site\cite{pachmari}
\item Time Structure: because the cosmic ray component contains penetrating particles
(mostly muons) that survive to detector level, the duration of the light pulse can be
longer. Many early versions of the ACT, particularly the Haleakala experiment\cite{haleakala},
attempted to exploit this feature but it was not to prove very effective,
\item  Spectral Content: the penetrating component of cosmic ray showers is close to
the light detector and its overall Cherenkov light at the detector is less attenuated
in the ultraviolet; this feature was used as a discriminant in the early
Whipple and Narrabri
experiments of Grindlay and his collaborators\cite{grindlay1975}
 and in the Crimean experiments\cite{crimea}. 
It is mostly effective when combined with other discriminants.
\item Angular Spread: the image of the light superimposed on the night-sky
background has a more regular distribution from gamma-ray showers and is smaller and more
uniform. This feature was recognized by Jelley and Porter\cite{jelleyporter1964} 
but not really exploited
until some decades later. This was to prove the most powerful discriminant and to lead to
the first successful credible detection of a TeV gamma-ray source\cite{weekes1989}.

\end{itemize}

The Cherenkov light image has a finite angular size which can, in principle,
be used to refine the arrival direction, and perhaps even to distinguish it from
the images of background cosmic rays\cite{jelley1965}.
However when a simple telescope with a single light  detector (pixel) is used as a
gamma-ray detector, this information is lost and the angular resolution
is no better than the field of view of the telescope.
Because the Cherenkov light images
are faint and fast, it is not technically straight-forward to record them.
 Boley and his collaborators \cite{boley} had used an array of photomultipliers
at Kitt Peak
to study the longitudinal development of large air showers but these were from
very energetic primaries. A pioneering effort by Hill and 
Porter\cite{hillporter1960}, using a image intensifier system
from a particle experiment, resulted in the first recorded images of Cherenkov
light from air showers (Figure~\ref{imaging}). These images, although relatively 
low resolution, demonstrated in a very vivid way the information contained
in the Cherenkov image recorded at ground level. The
potential advantages of using this detection technique
 as a means of separating out the gamma-ray
component were recognized in
a prophetic paper by John Jelley and Neil Porter\cite{jelleyporter1964}: 

``For a long time it has been appreciated that the image intensifier
offers potentialities in this field, and the photography of Cherenkov images 
against the night-sky is the first step in this direction. Temporarily
postponing the technical problems, what are the advantages of this technique? 
First, with Schmidt optics, it is possible in principle to combine a wide 
field of view with a high resolution. Secondly, photographs already obtained
of the Cherenkov images suggest that their shapes may be used to give detailed 
information both on the true direction of the shower and also the coordinates 
of its point of intersection with the ground, in relation to the position 
of the equipment. The third feature,
and it is really the most important one for gamma-ray astronomy of `point
sources', is the high angular resolution which may be attained. Though the 
Cherenkov images are $\approx$ 2$^\circ$ across, and are in general 
non-circular in shape, it should 
be possible to determine a shower direction to $\approx$ 0.2$^\circ$. Thus, we
have, for a true point source, a discrimination (by solid angle) against
showers from the general-field CR primaries, of $\approx$ 100 times
better than that possible for drift-scans with a photomultiplier system. 
It might be added
here that a stereoscopic technique, with two separated telescopes, would 
greatly enhance these potentialities.''

However, because of the finite size of the photocathode on the image intensifiers
then available,
it was only possible to couple them to a relatively small mirrors which meant
that only cosmic ray primaries above 100 TeV could be detected. Even then it
was necessary to couple these state-of-the-art instruments to a phosphor 
with decay times of microseconds to allow the image intensifier to be gated 
and the image recorded photographically.
Since this meant that the
technique was limited to energies $>$ 100 TeV where the attenuation of the 
gamma-ray flux by
photon-photon pair production in intragalactic space was appreciable, this
approach was not pursued at that time. A recent Japanese experiment has revived
interest in this technique using the best modern image intensifiers
(Sasaki, this workshop).

\begin{figure}
\begin{center}
\includegraphics[width=.7\textwidth]{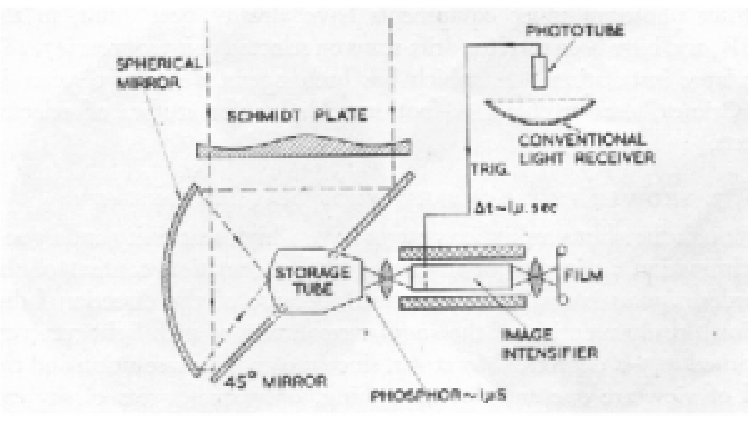}
\includegraphics[width=.7\textwidth]{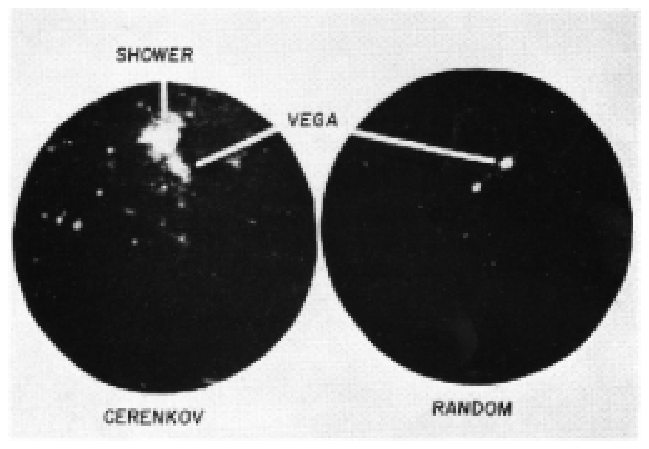}
\end{center}
\caption[]{{\bf Top:} Image Intensifier used by Hill and Porter to record 
the images of cosmic ray air showers \cite{hillporter1960}. {\bf Bottom}
Images of the night-sky triggered by an ACT (left) and triggered randomly
(right). The field of view was $\pm$12.5$^\circ$.}
\label{imaging}
\end{figure}

A novel approach to imaging  was that pursued by Grindlay and his colleagues
in the seventies\cite{grindlay1975} in
which multiple light detectors separated by distances $\approx$ 100 m
 were used to detect the shower maximum
associated with gamma-ray showers; this pinpointed the shower arrival direction.
The penetrating, mostly muon, component
from hadron showers was detected by a second detector and 
was used as a veto to preferentially select events
that were initiated by gamma rays.
This ``Double Beam'' technique was
potentially powerful but was difficult to implement with the resources 
available at the
time. Initially the detectors used were 1.5 m searchlight mirrors with single
phototubes at their foci; later the 10 m reflector was incorporated into the
system with two pixels. The technique received new life when the 
Narrabri Stellar Interferometer (in Australia) became
available. With two large reflectors of 9 m aperture on a circular rail system,
(Figure~\ref{double}) the system, originally built to measure the diameters
of bright stars using the intensity interferometer principle,
was ideally suited for this technique. Although some detections were reported 
(the Crab
pulsar, the Vela pulsar and Centaurus A)\cite{grindlay}, they were not 
confirmed
by later, more sensitive, observations. The Double-Beam technique, although
ingenious, was not pursued after this although it can be seen as the stalking horse
for imaging arrays (see below).

\begin{figure}
\begin{center}
\includegraphics[width=.6\textwidth]{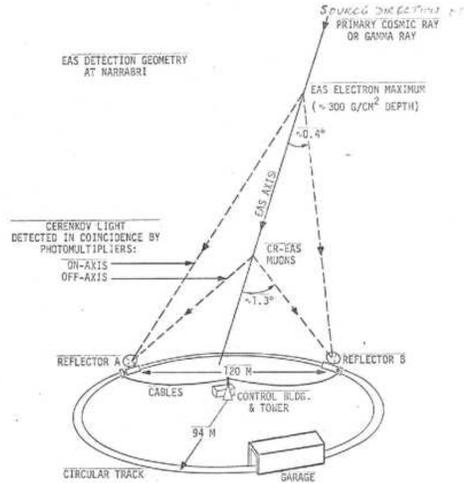}
\end{center}
\caption[]{The Double Beam Technique developed by Grindlay in which two reflectors
are used, each with two ``pixels.'' The upper two define the shower maximum and the lower
two define the penetrating component and act as a veto to reject hadronic showers.}
\label{double}
\end{figure}

Activity in ground-based gamma-ray astronomy was at a low ebb in the seventies.
Observations with the Whipple 10 m reflector had moved the energy threshold of the
technique close to 100 GeV but this had only produced upper limits on the predicted sources.
Smaller telescopes produced tentative detections of several binaries and pulsars but these
were always on the edge of statistical credibility and were not subsequently verified 
(this controversial epoch of TeV gamma-ray astronomy has been reviewed 
elsewhere\cite{chadwick},\cite{weekes1991}).

\subsection{The Power of the Atmospheric Cherenkov Imaging Technique}
The concept of using electronic cameras consisting of matrices
 of phototubes in the focal plane of large reflectors to record 
the images of the Cherenkov light from small air showers was first suggested in
 a paper at a workshop in Frascati, Italy\cite{weekesturver1977}. Entitled
``Gamma-Ray Astronomy from 10-100 GeV: a New Approach'' the emphasis was on lowering
the energy threshold through the use of two large reflectors separated by 100 m, each
equipped with arrays of phototubes in their focal plane. The motivation to go to lower
energies came from the prediction from Monte Carlo simulations that the ratio of Cherenkov
light from gamma-ray showers to cosmic ray showers of the same energy increases
dramatically below 100 GeV. In this paper the physical explanation of this falloff
was stated: ``In a proton shower most of the Cherenkov light comes from the 
secondary electromagnetic cascades.
Energy comes into these cascades via the production of pions by the primary 
and the subsequent nucleon cascade. Two thirds of the energy (approximately)
goes to charged pions; they can decay to muons or undergo a collision.The 
latter process is a more efficient method of producing Cherenkov light;
since the lifetime against decay is greater a higher energies, the chance 
of collisions is greater. At lower energies therefore, proportionally
more energy comes off in muons whose energy may be below the Cherenkov 
threshold and hence the low energy
showers are deficient in Cherenkov light''. 
The idea of using an array of phototubes with limited resolution to image the  
Cherenkov light rather than the high resolution offered by image intensifiers was 
motivated by the experience of the author using CCD detectors in optical astronomy 
where the resolution achieved is significantly greater than the scale of the
pixels. In the paper there was little emphasis on discrimination of the primaries based on
the shapes of the images although it was claimed that there would be a significant
improvement in angular resolution (to 0.25$^\circ$). The use of two reflectors in
coincidence was advocated to reduce the predicted muon background.

In this paper\cite{weekesturver1977} the basic concept of the Cherenkov light
 imaging telescope was described; it consisted of an array of PMTs in
the focal plane of a large reflector. Although the initial development 
centered on the use of a single large reflector (the Whipple 10 m reflector, 
Figure~\ref{whipple}), 
the utility of an array with at least two
such cameras was advocated. This has been the model for all subsequent
telescopes using the ACIT. In general,
in recording the Cherenkov light image from an air shower, the gamma-ray astronomer
tries to characterize its nature (gamma-ray or hadron), determines its arrival direction,
and gets some estimate of the primary that initiated the air shower.  
The factors that cause
the observed shape and size of the image are many: the nature of the primary particle, its
energy and trajectory, the physical processes in the particle cascade (principally pair
production and bremsstrahlung in electromagnetic cascades with the addition of pion
production in hadron initiated cascades), Coulomb scattering of shower electrons,
the effect of geomagnetic deflections of the
shower particles, the distance of the point of impact of the shower core from the
optic axis, the Cherenkov angle of emission, and the effect 
of atmospheric absorption\cite{fegan1997}.
In addition the properties of the imaging system must be completely understood: the
reflectivity of the mirrors, the quantum efficiency of the light detectors as a function
of wavelength, the time response of the system, and the distortions introduced by the
system's optics, cables, electronics and data readout.

Fortunately all of these factors are amenable to calculation or measurement. The physics
of the various processes involved in the shower development are well known and Monte
Carlo methods can be used to estimate the expected values from particular primaries.
However since fluctuations play a major role in such development the expected values
cover a range of possibilities and identification must always be a statistical process.
It is relatively easy to predict the properties of the gamma-ray initiated showers; it is
more difficult to predict the expected properties of the background which is mainly from
charged cosmic rays. While every attempt is made to estimate both signal and background,
it is usually found that the background contains some unpleasant surprises; hence although the
gamma-ray detection rate can be reliably predicted, the efficiency of the identification
of the gamma rays from the more numerous background requires the system to be actually
operated in observations of a known source. Since the background is numerous and
constant, its properties can be readily modeled from empirical databases 
of night-sky background events.
There is an irreducible background from hadron showers which develop like
electromagnetic cascades (most of the energy goes into a $\pi^o$ in the first interaction)
and from the electromagnetic cascades produced by cosmic electrons (whose fluxes in the
range of interest are 0.1 - 0.01\% of the hadron flux).

\subsection{The First Source}

When the imaging systems first went into operation it was not immediately obvious how
the images should be characterized and discriminated from the background. There were
no credible sources and Monte Carlo calculations were still being developed
and were untested.
The first such calculations available to the Whipple Collaboration indicated that
fluctuations might effectively rule out any discrimination and did not encourage
the development of sophisticated analysis techniques. The first Whipple camera had 37
pixels, each of 0.25$^\circ$ diameter\cite{cawley}. A relatively simple image parameter,
{\it Frac2}, defined as the ratio of the signal in the two brightest pixels to the
total light in the image, was developed empirically and led to the first indication
of a signal from the Crab Nebula\cite{cawley1985},\cite{gibbs}. This simple parameter
picked out the compact images expected from electromagnetic cascades but did not
provide any information on the arrival direction (other than that it was within
the field of view of the detector). However the application
of the same selection method on putative signals from the then popular sources, Cygnus
X-3 and Hercules X-1, did not improve the detection credibility and initially cast doubt
on the effectiveness of {\it Frac2} as a gamma-ray identifier.

Since the images were roughly elliptical in shape, an attempt was made to quantify the
images in terms of their second and third moments\cite{mckeown1983}. However this was not
applied to gamma-ray identification until Hillas  undertook a new series of Monte
Carlo calculations\cite{hillas1985}.
These calculations predicted that gamma-rays images could be
distinguished from the background of isotropic hadronic images based on two criteria:
the difference in the {\it physics}
 of the shower development, which led to smaller and better defined
ellipses for gamma rays, and the difference in the {\it geometry} of image formation
due to all images coming from a point source on axis having their major axes
intersecting the center of the field of view.
Fortunately the first property aids the definition of the second and provides potentially
very good angular resolution. Hillas\cite{hillas1985} defined  a series of parameters
which included the second moments ({\it Width} and {\it Length}), the parameter
{\it Dist} which measures the distance of the centroid of the image from the optic axis,
and {\it Azwidth} which measures the projected width of the image on the line joining the
centroid to the center of the field of view. Later {\it Alpha}, the angle between this
line and the major axis was added as was {\it Asymmetry}, the third moment. {\it Azwidth}
was particularly simple; it is easy to use and proved to be very effective as it combined
discrimination based on image size (physics) and arrival direction (geometry) and led
   to the first definite detection of a point source of TeV gamma-rays. In general
multiple parameter selections were made. The
parameters were first defined in Monte Carlo calculations but once the standard
 candle
of the Crab Nebula was established\cite{weekes1989}, optimization was made on
the strong and steady Crab
signal to preferentially select gamma rays. This optimization led to an 
analysis package
called Supercuts\cite{punch}, which proved to be extraordinarily robust, and 
in various
forms, was the basis of the data analysis used by the Whipple Collaboration to
detect the first AGN\cite{punch1992},\cite{quinn1996},
\cite{holder},\cite{catanese},\cite{horan2003}.
Other groups have defined different
parameters and analysis schemes but the basic methodology is the same.

\section{ACT Observatories}

\subsection{Third Generation Observatories}

By 1996 the ACIT was judged to have been very successful and a number of groups
made plans for third generation ACTs. The limitation of a single telescope
was easily seen from the results obtained using the Whipple telescope 
and camera\cite{kildea05}. At low trigger thresholds it was impossible to distinguish low
energy gamma-ray events from the much more numerous background of partial muon
rings (arcs). Despite intense efforts with sophisticated analysis methods, it was
clear that the discrimination threshold was a factor of 2-3 above the trigger
threshold. Hence although the fundamental threshold was $\approx$ 200 GeV, the effective
gamma-ray threshold was $\approx$ 400 GeV. 
Since the muon Cherenkov emission is essentially a
local phenomenon, this background is easily eliminated by demanding a
coincidence with a second telescope separated from the first by a minimum
distance of 50 m\cite{weekesturver1977}. In fact the HEGRA experiment had
already demonstrated\cite{hegra} the power of an array of small imaging telescopes
to improve the angular and energy resolution of the ACIT; at the threshold
energies of these telescopes the muon background was not a problem.

Thus it was apparent that the next generation of the ACIT would involve arrays of reflectors
with apertures in excess of 10 m, with better optics, with more sophisticated
cameras, and with data acquisition systems capable of handling high rates. Such
systems required an investment that was almost an order of magnitude greater
than the previous generation of detectors (but the flux sensitivity would be
improved by a similar factor). Of necessity the number
of people involved in each experiment would be so large ($\approx$ 100)
that the new collaborations
would be more in line with the numbers of scientists found in particle physics
experiments than in typical major astronomical projects.

\subsection{The Power of ACT Arrays}

ACTs arrays can be discussed under the headings of improvements offered in energy threshold,
energy resolution, angular resolution and background discrimination. A comprehensive discussion
can be found in\cite{Aharonian:97}. A typical array provides multiple images of
a single event as seen in Figure~\ref{cartoon}.

\begin{figure}
\begin{center}
\includegraphics[width=.8\textwidth]{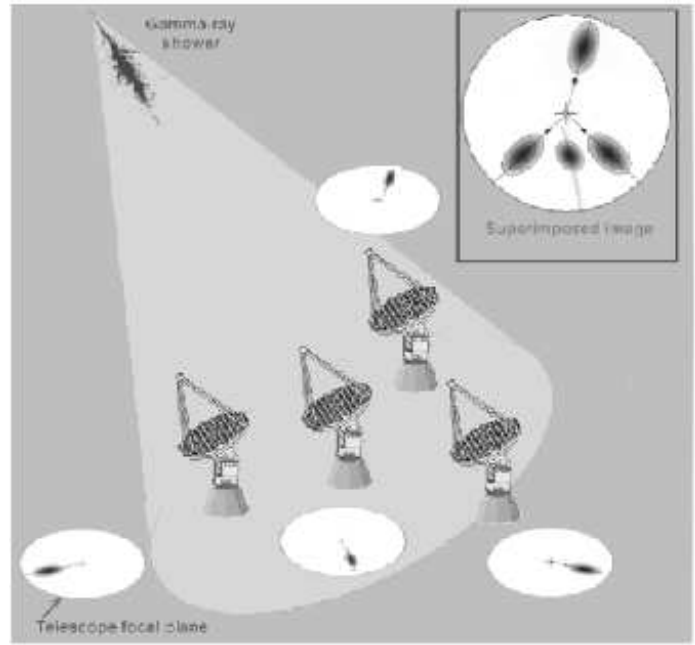}
\end{center}
\caption[]{Cartoon showing response of array of four detectors to
air shower  whose axis is parallel to the optical axes of the
telescopes and some 30 m displaced from the center of array.
(Figure courtesy of P.Cogan)}
\label{cartoon}
\end{figure}

\noindent
{\bf Energy Threshold:}
The basic quantities involved in determining the energy threshold of an ACT
are given above in Section 2.3 and are fairly obvious: the mirror area should be as
large as possible and the light detectors should have the highest possible
quantum efficiency. To the first approximation (as demonstrated in\cite{chudakov1965})
it does not critically depend on how the mirror area is distributed, i.e., a cluster
of small telescopes in close proximity operated in coincidence is the same as if their
signals are added and is approximately the same as that of a single large telescope
of the same total mirror area. Practical considerations tend to dominate: coincidence
systems are more stable, the cost of telescopes scales as the Aperture$^{2.5}$, the
relative cost of multiple cameras each on a small telescope versus the
cost of a single camera on a large telescope, etc. However the simplest way to get
the lowest energy threshold is to go for a single large telescope (although this
may introduce other problems).

\noindent
 {\bf Angular Resolution:}
Angular resolution is important not only for reducing the background and identifying
a potential source but also for mapping the distribution of gamma rays in the
source.
Stereoscopic imaging, the simplest form of ``array'' imaging, offers the immediate
advantage of improving the angular resolution. This principle was established
with the use of just two telescopes with a separation of $\approx$ 100 m, i.e., with the
two telescopes within the light pool of the Cherenkov light pool, $\approx$
a circle of diameter 200 m. The greater the separation, the better the
angular resolution but increasing the separation beyond 100 m begins to reduce
the effective gamma-ray collection area. A simple array of imaging
ACTs can provide a
source location of $\approx$ 0.05$^\circ$ for a relatively strong source with angular
resolution of $\approx$ 0.1$^\circ$ for individual events. This is a factor of two
improvement over that for a single telescope.
An angular resolution of an arc-min or better appears feasible ultimately.

\noindent
{\bf Background Discrimination:}
Multiple views of the same air shower from different angles obviously improves
the signal-to-noise ratio when the images are combined. However in reducing the
background of hadronic events the gain is not as large as might appear at
first glance. Hadronic showers which develop like typical showers are
easily identified and rejected, even in a single telescope. More subtle are
the hadronic events which develop like an electromagnetic cascade (an
early interaction channels much of the energy into an electron or
gamma ray). Such events cannot be identified no matter how many views are
provided on the cascade development. Similarly the cascades initiated by
cosmic electrons are an irreducible background. However the array approach
does completely remove the background from single local muons and the
improved angular resolution narrows the acceptable arrival directions.

\noindent
{\bf Energy Resolution;}
The Cherenkov light emitted from the electromagnetic cascade is to a first
approximation proportional to the energy of the initiating gamma ray
and thus can be considered a calorimetric component. However
with a single ACT there is no precise information as to the impact parameter
of the shower axis at ground level. Since the intensity of the Cherenkov light
is a function of distance from the shower axis, the lack of information on this
parameter is the limiting factor in determining the energy of the gamma ray. The
energy resolution of a single imaging ACT is $\approx$ 30-40\%.
   With an array the impact parameter can be determined to $\approx$ 10 m and the energy
resolution, in principle, can be reduced to 10\%.

\subsection{The Third Generation Arrays}

This third generation of ACTs has seen the formation of four
large collaborations formed to build  arrays of large telescopes:
a largely German-Spanish
collaboration that is building two 17 m telescopes on La Palma in
the Canary islands (MAGIC)\cite{magic}: an
Irish-British-Canadian-USA collaboration that is building an array of four
12 m telescopes in Arizona (VERITAS)\cite{veritas}; an Australian-Japanese
collaboration that has built four 10 m telescopes in Australia
(CANGAROO-III)\cite{cangarooIII}; a largely European collaboration that has built
an array of four 12 m telescopes in Namibia (HESS)\cite{hess}
and plans to add a fifth
telescope of 28 m aperture at the center of the array. 
The fact that two of the arrays are in each hemisphere
is somewhat fortuitous but ensures that there will be good coverage of the
entire sky and that all observations can be independently verified.
Three of arrays are discussed elsewhere at this workshop; here the VERITAS 
observatory will be briefly described.

The sensitivity of these new arrays is probably not dissimilar; HESS and MAGIC
has demonstrated what can achieved in the actual detection of known and new sources.
With the second generation of ACTs (Whipple, HEGRA), it was possible to detect
a source that was 5\% of the Crab Nebula in 100 hours of observation. With
HESS this is reduced to one hour and in principle in 100 hours it should be
possible to detect a source as weak as 0.5\% of the Crab. HESS has also
demonstrated an energy resolution of 10\% and an angular resolution of
an arc-min.

\section{VERITAS}
The configuration chosen for VERITAS was 
a filled hexagon of side 80 m; in the first phase funding was available for 
only four telescopes so the hexagon has
three non-adjacent vertices missing\cite{veritas}.
The four telescopes and cameras of VERITAS are identical and are now
at an advanced state of construction.
The first two telescopes and cameras were installed
at a temporary site (the Whipple Observatory Basecamp at an
elevation of 1.3 km) and
saw first ``gamma-ray light'' in February, 2005
(Figure~\ref{Tel1}). The properties of the first telescope have been
described elsewhere\cite{Weekes05},\cite{holder06}.

{\bf Telescope:}
The VERITAS telescopes are of the Davies-Cotton optical design with 12 m
aperture and 12 m focal length.
The mechanical structure consists of an
altitude-azimuth positioner and a tubular steel optical support
structure (OSS).
The design is closely modeled on the existing
Whipple 10 m optical reflector but with the added feature of a mechanical
bypass of the upper quadrapode arm which transfers the load of the
camera to the counterweight support.
Completion of the first two telescopes has allowed the properties and
sensitivities of the individual telescopes to be measured\cite{holder06}.

The 350 individual mirror facets on each telescope are hexagonal, each with an area of
0.322 m$^{2}$, providing a total mirror area of $\sim$110 m$^{2}$. They
are made from glass, slumped and polished;
the glass facets are
aluminized and anodized at the VERITAS optical coating
laboratory on-site.
The reflectivity of the anodized coating is typically $>90\%$ at
320 nm. Each facet has a 24 m radius of curvature. They are located
on a three point mounting on the spherical front surface
(radius 12 m) of the OSS. The point
spread function (PSF) at the position of Polaris (elevation ~ 31$^\circ$)
        was measured to be
$0.06^\circ$ FWHM; with bias alignment it is anticipated that this
PSF will be achieved over most of the operating
range of VERITAS.

\begin{figure}[htb]
\vspace*{-1.0in}
%\begin{tabular}{c}
\hspace*{-0.15in}
{\rotatebox{270}{\includegraphics*[height=12cm]{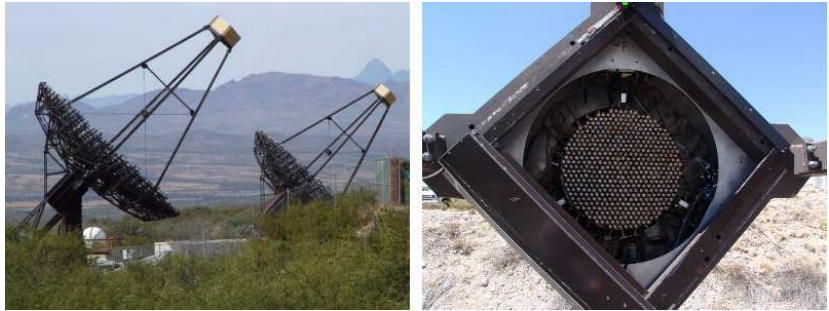}}}
%\hspace{0.5cm}
%\includegraphics*[height=6cm]{weekes_tokyo_fig8bb.ps}
%   \end{tabular}
\vspace*{-1.0in}
    \caption{\label{Tel1}
      {\bf Left:} The first two VERITAS Telescopes. {\bf Right:} The 499 pixel PMT camera.
    }
%\end{center}
\vspace{-0.3cm}
\end{figure}

{\bf Camera:}
The VERITAS cameras are closely modeled on those used previously by the
group at the Whipple telescope but incorporate much more advanced triggering,
electronics readout and data acquisition systems \cite{gibbs}.
The instrumentation in the focal plane is a 499 element photomultiplier tube (PMT)
camera, with $0.15^\circ$ angular spacing giving a field-of-view of
$3.5^\circ$. The camera is shown in Figure~\ref{Tel1}.
 The PMTs are Photonis XP2970/02 with a quantum efficiency
$>20\%$ at 300 nm, currently operated at a gain of
$\sim2\times10^{5}$.
The PMT signals are amplified by high bandwidth preamplifiers integrated into
the PMT base mounts.
The signals are sent via $\sim$50 m
of RG59 stranded cable to the telescope trigger and data acquisition
electronics, at which point the observed pulse for an input delta function has a
rise time (10\% to 90\%) of 3.3 ns and a width of 6.5 ns.

The PMT signals are digitized using custom-built VME boards housing Flash ADCs
with 2 ns sampling and a memory depth of $32\mu$s.
The trigger system is multi-level. At the telescope each channel is equipped
with a programmable constant fraction
discriminator (CFD) for each PMT, the output of which is passed
to a pattern recognition trigger system which is
programmed to recognize triggers resembling true compact Cherenkov light
flashes. Individual telescope triggers are delayed and combined to form
an overall array trigger. The FADCs permit the telescopes to operate at a
lower threshold than would otherwise have been possible\cite{holder06}.

{\bf Future Program:}
The scientific program will concentrate on the study of
extragalactic objects including AGN, radio galaxies, starburst galaxies and
clusters, compact galactic objects including pulsars, binaries and
microquasars, extended objects such as supernovae remnants, unidentified
sources discovered in future space missions, and signatures of dark
matter in the center of galaxies. A sky survey will be undertaken and the
study of gamma-ray bursts will have high priority.

{\bf Acknowledgments:} Over the past 40 years ground-based gamma-ray astronomy
at the Smithsonian's Whipple Observatory has been supported by times by the
Smithsonian Astrophysical Observatory, the U.S. Department of Energy, the National
Science Foundation and NASA. D.J Fegan is thanked for helpful comments on 
the manuscript.

\end{document}